\documentclass[11pt,a4paper,fleqn]{article}

\usepackage{amsmath,amssymb,amsthm,enumerate,cite}

\newcommand{\vspacebefore}{\raisebox{0ex}[2.5ex][0ex]{\null}}

\newcommand{\p}{\partial}

\newcommand{\const}{\mathop{\rm const}\nolimits}
\newcommand{\Equiv}{\mathop{\sim}}
\newcommand{\pot}{\mathop{\rm pot}}

\newcommand{\thetbn}{\arabic{nomer}}

\newtheorem{theorem}{Theorem}
\newtheorem{lemma}{Lemma}
\newtheorem{corollary}{Corollary}

{\theoremstyle{definition} \newtheorem{definition}{Definition}

\newtheorem{note}{Note}
\begin{document}

\par\noindent {\LARGE\bf Potential equivalence transformations \\ for nonlinear
diffusion--convection equations\par}
{\vspace{4mm}\par\noindent {\it Roman~O.~Popovych~$^\dag$ and Nataliya~M.~Ivanova~$^\ddag$
} \par\vspace{2mm}\par}
{\vspace{2mm}\par\noindent {\it
Institute of Mathematics of National Academy of Sciences of Ukraine, \\
3, Tereshchenkivska Str., Kyiv-4, 01601, Ukraine\\
}}
{\noindent {\it
$^\dag$rop@imath.kiev.ua, $^\ddag$ivanova@imath.kiev.ua
} \par}

{\vspace{5mm}\par\noindent\hspace*{8mm}\parbox{140mm}{\small
Potential equivalence transformations (PETs) are effectively applied
to a class of nonlinear diffusion--convection equations.
For this class  all possible potential symmetries are classified and
a theorem on connection of them with point ones via PETs is also proved.
It is shown that the known non-local transformations between equations
under consideration are nothing but PETs.
Action of PETs on sets of exact solutions of a fast diffusion equation 
is investigated.
}\par\vspace{5mm}}

\section{Introduction}

In this paper we consider a class of nonlinear diffusion--convection equations of the form
\begin{equation} \label{eqf1}
u_t=(d(u)u_x)_x+k(u)u_x,
\end{equation}
that have a number of applications in mathematical physics
(see for instance~\cite{Barenblatt1,Barenblatt2,Kamin82,Philip1988}).
Equation~\eqref{eqf1} is called also the Richard's one~\cite{Richard's1931,Wiltshire&El-Kafri2004}.

Here $d=d(u)$ and $k=k(u)$ are arbitrary smooth functions of $u,$ $d(u)\!\neq\! 0.$
The linear case of~(\ref{eqf1}) ($d,k=\const$) was studied by S.~Lie~\cite{lie1881}
in his classification of linear second-order PDEs with two independent variables.
(See also modern treatment of this subject in~\cite{Ovsiannikov1}.)

Various classes of quasi-linear evolutionary equations in two independent variables
that intersect class~(\ref{eqf1}) were investigated by means of symmetry methods
in~\cite{Basarab-Horwath&Lahno&Zhdanov2001,Cherniha&Serov,Edwards,Katkov1968,
Wiltshire&El-Kafri2004,Sposito1990,Oron&Rosenau,Ovsiannikov1959,Yung&Verburg&Baveye1994}.
The~complete and strong group classification of~(\ref{eqf1})
as well as a review of previous results on this subject
were presented in~\cite{Popovych&Ivanova2003NVCDCEsLanl}.

\looseness=-1
To study non-local symmetries of PDEs in the framework of the local approach,
Bluman {\it et al} \cite{Bluman88,Bluman89} proposed the notion of
{\it potential symmetries}.
A system of PDEs may admit symmetries of such sort
when some of the equations
can be written in a conserved form.
After introducing potentials for PDEs written in the conserved form as
additional dependent variables, we obtain a new (potential) system of PDEs.
Any local invariance transformation of the obtained system induces
a symmetry of the initial system.
If transformations of some of the ``non-potential" variables
explicitly depend on potentials, this symmetry
is a non-local (potential)
symmetry of the initial system.
More details about potential symmetries and their applications
can be found in~\cite{Bluman88,Bluman89,bluman93}.
Potential symmetries of~(\ref{eqf1}) and its generalizations were studied
by C.~Sophocleous~\cite{Sophocleous1996,Sophocleous2000,Sophocleous2003}.
Other approaches for investigation of non-local symmetries of~(\ref{eqf1}) were used
in~\cite{Akhatov&Gazizov&Ibragimov,Pukhnachov1996}.
I.G.~Lisle~\cite{LisleDissertation} obtained a number of results concerning
equivalence transformations of~(\ref{eqf1}) and~(\ref{pot1f1})
and group classification in these classes.
Unfortunately, these results, including the notion of potential equivalence
transformations (PETs), were little known until now and were rediscovered by other scientists.

In this paper we study in detail connections between symmetries and equivalence transformations
of equations~(\ref{eqf1}), corresponding potential systems and equations for the potential.
For class~(\ref{eqf1}) we prove a theorem on connection of the potential symmetries
with local ones via PETs.
It is shown that all the known non-local transformations between equations
from the class under consideration are, in fact, PETs.
In particular, they include the well-known transformations linearizing
the $u^{-2}$-diffusion (named also Fujita--Storm) equation~\cite{Bluman&Kumei1980,Storm1951},
Fokas--Yortsos~\cite{Fokas&Yortsos1982,Strampp1982b} and Burgers equations~\cite{Cole1951,Forsyth1906,Hopf1950}
as well as the less known transformation of ``logarithmic nonlinearity" to ``power nonlinearity".
For some equations PETs are nonlocal symmetries. In such cases they generate additional equivalences 
on the corresponding sets of solutions and can be used e.g. to construct new exact solutions from known ones. 

Our paper is organized as follows. 
In Section~2 known results on classical group analysis of diffusion--convection equations are adduced in 
a form which is suitable for purposes of our investigations. 
After formulating the statement of problem on classification of potential symmetries rigorously, 
we completely classify symmetries of such kind for the equations under consideration in Section~3.
Analysis of connections between potential and Lie symmetries of equations from class~(\ref{eqf1}),
which is given in Section~4, is essentially based on accuracy of the above classification results.  
The main theorem on reducibility of potential symmetries to point ones with potential and additional point 
equivalence transformations is proved for class~(\ref{eqf1}). 
Section~5 is devoted to demonstration of using PETs as non-local invariance transformations. 
As an example, the fast diffusion equation~$u_t=(\ln u)_{xx}$ is considered in such framework.

\section{Group classification of diffusion--convection equations}

The exhaustive result on classical group classification of class~(\ref{eqf1})
is presented by the statements adduced below~\cite{Popovych&Ivanova2003NVCDCEsLanl}.

Using the direct method, we construct the complete equivalence group including 
both continuous and discrete point transformations.

\begin{theorem}
Any transformation from the equivalence group $G^{\Equiv}$ has the form
\[
\tilde t=\varepsilon_4t+\varepsilon_1, \quad
\tilde x=\varepsilon_5x+\varepsilon_7 t+\varepsilon_2, \quad
\tilde u=\varepsilon_6u+\varepsilon_3, \quad
\tilde d=\varepsilon_4^{-1}\varepsilon_5^2d, \quad
\tilde k=\varepsilon_4^{-1}\varepsilon_5k-\varepsilon_7,
\]
where $\varepsilon_1,$ \dots, $\varepsilon_7$ are arbitrary constants,
$\varepsilon_4\varepsilon_5\varepsilon_6\ne0.$
\end{theorem}

\begin{corollary}
If the equations from class~(\ref{eqf1}) are rewritten in the explicit conserved form
\begin{equation} \label{eqf1ecf}
u_t=(d(u)u_x-K(u))_x
\end{equation}
where $K_u=-k,$ then the corresponding equivalence group~$\widehat G^{\Equiv}$ consists of the transformations
\begin{gather*}
\tilde t=\varepsilon_4t+\varepsilon_1, \quad
\tilde x=\varepsilon_5x+\varepsilon_7 t+\varepsilon_2, \quad
\tilde u=\varepsilon_6u+\varepsilon_3, \\
\tilde d=\varepsilon_4^{-1}\varepsilon_5^2d, \quad
\tilde K=\varepsilon_4^{-1}\varepsilon_5\varepsilon_6K+\varepsilon_7 u +\varepsilon_8,
\end{gather*}
where $\varepsilon_1,$ \dots, $\varepsilon_8$ are arbitrary constants,
$\varepsilon_4\varepsilon_5\varepsilon_6\ne0.$
\end{corollary}

\begin{theorem}
The Lie algebra of the kernel of principal groups of~\eqref{eqf1} is
$A^{\ker}=\langle \p_t,\; \p_x \rangle.$
A~complete set of $G^{\Equiv}$-inequivalent
equations~\eqref{eqf1} with the maximal Lie invariance algebra $A^{\max}$
not equal to $A^{\ker}$
is exhausted by cases given in Table~1.
\end{theorem}

\begin{note}
In Table~1 $\mu,\nu=\const$. $(\mu,\nu)\not=(-2,-2),\,(0,1)$ and
$\nu\not=0$ for Case~1.\ref{t56}. $\mu\not=-4/3,0$ for Case~1.\ref{t57}a.
The function $h=h(t,x)$ is an arbitrary solution
of the linear heat equation ($h_t=h_{xx}$).
Case~1.\ref{t57}b can be reduced to~1.\ref{t57}a ($\mu=-2$) by means of the additional
equivalence transformation
\begin{equation}\label{AdEqTr}
\tilde t=t,\qquad \tilde x=e^x,\qquad \tilde u=e^{-x}u.
\end{equation}
\end{note}

\begin{note}
Hereafter for convenience we use double numeration T.N of classification cases,
where T denotes the number of table and N denotes the number of the respective case in Table~T.
The notion ``equation~1.N'' ("system~2.N'') is used for the equation of form~\eqref{eqf1}
(the system of form~\eqref{pot1f1}) where
the parameter-functions take values from the corresponding case.
\end{note}

\vspace{-2ex}

{ 
\newcounter{tbn} \setcounter{tbn}{-1}
\begin{center}
{\bf Table~1.} Results of group classification for class~(\ref{eqf1})\\[1ex]
\footnotesize\renewcommand{\arraystretch}{1.2}
\begin{tabular}{|l|c|c|l|}
\hline\vspacebefore
N &$d(u)$ & $k(u)$ & \hfill Basis of $A^{\max}\hfill$ \\
\hline\vspacebefore
\refstepcounter{tbn}\label{t51}\thetbn&$\forall$ & $\forall$ & $\p_t,\;\p_x  $
\\
\refstepcounter{tbn}\label{t52}\thetbn&$\forall$ &0 &$\p_t,\;\p_x,\;2t\p_t+x\p_x$
\\
\refstepcounter{tbn}\label{t53}\thetbn&$e^{\mu u}$ &$e^u$ &$\p_t,\;\p_x,\;
                             (\mu-2)t\p_t+(\mu-1)x\p_x+\p_u$
\\
\refstepcounter{tbn}\label{t54}\thetbn&$e^u$ & $u$ &
   $ \p_t,\;\p_x,\;t\p_t+(x-t)\p_x+\p_u$
\\
\refstepcounter{tbn}\label{t55}\thetbn&$e^u$ & 0 &
   $ \p_t,\;\p_x,\;2t\p_t+x\p_x,\;t\p_t-\p_u$
\\
\refstepcounter{tbn}\label{t56}\thetbn&$u^{\mu}$ & $u^\nu$ &
   $ \p_t,\;\p_x,\;(\mu -2\nu)t\p_t+(\mu -\nu)x\p_x+u\p_u$
\\
\refstepcounter{tbn}\label{t59}\thetbn&$u^\mu$ & $\ln u$ &
   $ \p_t,\;\p_x,\;\mu t\p_t+(\mu x-t)\p_x+u\p_u$
\\
\refstepcounter{tbn}\label{t57}\thetbn a&$u^{\mu}$ & 0 &
   $ \p_t,\;\p_x,\;2t\p_t+x\p_x,\;\mu t\p_t-u\p_u$
\\
   \thetbn b&$u^{-2}$ & $u^{-2}$ &$ \p_t,\;\p_x,\;2t\p_t+u\p_u,\;e^{-x}(\p_x+u\p_u)$
\\
 \refstepcounter{tbn}\label{t58}\thetbn&$u^{-4/3}$ & 0 &
   $ \p_t,\;\p_x,\;2t\p_t+x\p_x,\;4t\p_t+3u\p_u,\;x^2\p_x-3xu\p_u$
\\
\refstepcounter{tbn}\label{t510}\thetbn&1 & $u$ &
$\p_t,\; \p_x,\; t\p_x-\p_u,\; 2t\p_t+x\p_x-u\p_u,\; t^2\p_t+tx\p_x-(tu+x)\p_u$
\\
\refstepcounter{tbn}\label{t511}\thetbn&1 & 0 &
$\p_t,\; \p_x,\; 2t\p_t+x\p_x,\; 2t\p_x-xu\p_u,\; 4t^2\p_t+4tx\p_x-(x^2+t)u\p_u,\; u\p_u,\; h\p_u$
\\
\hline
\end{tabular}
\end{center}

\begin{note}
The exponential cases~1.\ref{t53}--1.\ref{t55} can be regarded as limits of
the power cases~1.\ref{t56}--1.\ref{t57}a.
More exactly,
\begin{gather*}
\tilde u=1+\nu^{-1}u,\; \mu=\mu'\nu\,{:} \quad
1.\ref{t56}_{\mu,\nu}\to 1.\ref{t53}_{\mu'},\; \nu\to+\infty,\\
\tilde u=1+\mu^{-1}u,\; \tilde t=\mu^2 t,\; \tilde x=\mu x\,{:} \quad
1.\ref{t59}_{\mu}\to 1.\ref{t54},\; \mu\to+\infty,\\
\tilde u=1+\mu^{-1}u\,{:} \quad
1.\ref{t57}{\rm a}_{\mu}\to 1.\ref{t55},\; \mu\to+\infty.
\end{gather*}
The above limits are extended to structure of the Lie invariance algebras and
can be used to obtain exact solutions for the exponential cases from ones for the power cases. 
Some partial cases of the above limits for diffusion equations ($k=0$) were adduced 
in~\cite{Bluman88,Bluman89}.
\end{note}

\section{Classification of potential symmetries}

After rewriting equation~(\ref{eqf1}) in the conserved form $u_t=(du_x-K)_x$ where $K'=-k$
and introducing the new (potential) unknown function $v=v(t,x)$,
we obtain the equivalent system of PDEs (called {\em potential} one)
\begin{equation}\label{pot1f1}
v_x=u, \quad v_t=du_x-K.
\end{equation}

It follows from system~(\ref{pot1f1}) that the function $v$ satisfies the equation
\begin{equation}\label{pot1f1a}
v_t=d(v_x)v_{xx}-K(v_x)
\end{equation}
that is called {\em potential equation} corresponding to equation~(\ref{eqf1}).
System~(\ref{pot1f1}) can be regarded as a Lie--B\"acklund transformation between
equations~(\ref{eqf1}) and~(\ref{pot1f1a}).

I.G.~Lisle proved in~\cite{LisleDissertation} that 
the Lie algebra of the equivalence group $G^{\Equiv}_{\pot}$ for the class of systems~(\ref{pot1f1}) is
\[\arraycolsep=0ex\begin{array}{ll}
A^{\Equiv}_{\pot}=\langle &\p_t ,\; \p_x,\; \p_u+x\p_v,\; \p_v,\;
t\p_t-d\p_d-K\p_K,\; x\p_x+v\p_v+2d\p_d+K\p_K,
\\[1ex]
&u\p_u+v\p_v+K\p_K,\; t\p_x+u\p_K,\; t\p_v-\p_K,\; v\p_x-u^2\p_u+2ud\p_d-uK\p_K \rangle.
\end{array}\]
He also constructed the connected component of unity in~$G^{\Equiv}_{\pot}$ and 
attached some discrete equivalence transformations to it. We prove using the direct method that 
the transformation group obtained in such way coincides with the whole equivalence group~$G^{\Equiv}_{\pot}$.

\begin{theorem}
Any transformation from $G^{\Equiv}_{\pot}$ has the form
\begin{gather*}
\tilde t=\varepsilon_1t+\varepsilon_2,\quad
\tilde x=\varepsilon_1'x+\varepsilon_2'v+\varepsilon_3't+\varepsilon_4',\quad
\tilde v=\varepsilon_1''x+\varepsilon_2''v+\varepsilon_3''t+\varepsilon_4'',
\\[1ex]
\tilde u=\dfrac{\varepsilon_1''+\varepsilon_2''u}{\varepsilon_1'+\varepsilon_2'u},\quad
\tilde d=\dfrac{(\varepsilon_1'+\varepsilon_2'u)^2}{\varepsilon_1}\,d,\quad
\tilde K=\dfrac{\varepsilon_1'\varepsilon_2''-\varepsilon_2'\varepsilon_1''}
{\varepsilon_1'+\varepsilon_2'u}\,\dfrac {K}{\varepsilon_1}
-\dfrac{\varepsilon_3''}{\varepsilon_1}
+\dfrac{\varepsilon_3'}{\varepsilon_1}\,
\dfrac{\varepsilon_1''+\varepsilon_2''u}{\varepsilon_1'+\varepsilon_2'u},
\end{gather*}
where $\varepsilon_1,$ $\varepsilon_2,$ $\varepsilon_i',$ $\varepsilon_i''$
$(i=\overline{1,4})$ are arbitrary constants,
$\varepsilon_1(\varepsilon_1'\varepsilon_2''-\varepsilon_2'\varepsilon_1'')\ne0.$
\end{theorem}

\begin{definition}
We call the transformations from $G^{\Equiv}_{\pot}$
{\it potential equivalence transformations} (PETs)
for class~\eqref{eqf1} (or~\eqref{eqf1ecf}).
\end{definition}

\begin{theorem}
The set~$G^{\Equiv}_{{\rm triv.}\pot}$ of potential equivalence transformations
which act on the arbitrary elements $d$ and $K$
trivially modulo~$\widehat G^{\Equiv}$
is a normal subgroup of~$G^{\Equiv}_{\pot}$.
The corresponding factor-group can be identified with the group
formed by the transformations
\begin{equation}\label{pottr}
\tilde t=t,\quad
\tilde x=x+\varepsilon v,\quad
\tilde u=\dfrac u{1+\varepsilon u},\quad
\tilde v=v,\quad
\tilde d=(1+\varepsilon u)^2d,\quad
\tilde K=\dfrac K{1+\varepsilon u},
\end{equation}
where $\varepsilon$ is an arbitrary real, and the hodograph transformation of variables $x$ and $v$
\begin{equation}\label{pothodograph}
\tilde t=t,\quad
\tilde x=v,\quad
\tilde u=u^{-1},\quad
\tilde v=x, \qquad
\tilde d=u^2d, \qquad
\tilde K=-u^{-1}K.
\end{equation}
\end{theorem}

\begin{definition}
We will call~\eqref{pottr} and~\eqref{pothodograph}
{\em purely potential} equivalence transformations for the class of PDEs~\eqref{eqf1}.
\end{definition}

Studying potential symmetries of~(\ref{eqf1}) is equivalent to
solving of the group classification problem in class of systems~(\ref{pot1f1})
with respect to the (incomplete) equivalence group~$G^{\Equiv}_{{\rm triv.}\pot}$.
Let us note that potential symmetries of~(\ref{eqf1}) were investigated
in~\cite{Sophocleous1996}.
Using transformations from $G^{\Equiv}_{{\rm triv.}\pot}$,
we essentially simplify, order and complete these results. 

\begin{theorem}
The Lie algebra of the kernel of principal groups of~\eqref{pot1f1} is
$A^{\ker}_{\pot}=\langle \p_t,\; \p_x,\; \p_v\rangle$
($=A^{\max}_{2.0}$).
A~complete set of $G^{\Equiv}_{{\rm triv.}\pot}$-inequivalent
systems~\eqref{pot1f1}
with the maximal Lie invariance algebra $A^{\max}$ not equal to
$A^{\ker}_{\pot}$ is exhausted by cases given in Table~2.
\end{theorem}

To test some results presented in Table~2, we used the unique program LIE by A.~Head~\cite{Head}.

\section{Analysis of classification results}

Let us analyze connections between cases from Tables~1 and~2.

Cases~2.\ref{t.ps.0z}$^*$--2.\ref{t.ps.8z}$^*$ completely correspond to Cases~1.\ref{t51}--1.\ref{t57}a:
2.\ref{t.ps.0z}$^*$$\leftrightarrow$1.\ref{t51},
2.\ref{t.ps.1z}$^*$$\leftrightarrow$1.\ref{t52},
2.\ref{t.ps.2z}$^*$$\leftrightarrow$1.\ref{t53},
2.\ref{t.ps.3z}$^*$$\leftrightarrow$1.\ref{t54},
2.\ref{t.ps.4z}$^*$$\leftrightarrow$1.\ref{t55},
2.\ref{t.ps.5z}$^*$$\leftrightarrow$1.\ref{t56}$_{\mu\not=-1}$,
2.\ref{t.ps.6z}$^*$$\leftrightarrow$1.\ref{t56}$_{\mu=-1}$,
2.\ref{t.ps.7z}$^*$$\leftrightarrow$1.\ref{t59},
2.\ref{t.ps.8z}$^*_{\mu\not=-4/3}$$\leftrightarrow$1.\ref{t57}a$_{\mu\not=-2}$.
The constant multiplier in $K$ ($\sim k$) can be change using equivalence transformations of the form
$\tilde t=\varepsilon^2 t,$ $\tilde x=\varepsilon x,$ $\tilde u=u,$ $\tilde v=\varepsilon v,$
$\tilde d=d,$ $\tilde K=\varepsilon K$ ($\sim\tilde k=\varepsilon k$), which nontrivially act only
on the latter basis operators in Cases~2.\ref{t.ps.3z}$^*$ and 1.\ref{t54}.
The correspondence 2.\ref{t.ps.7z}$^*$$\rightarrow$1.\ref{t59} are established with
the equivalence transformation $\tilde t=t,$ $\tilde x=-x,$ $\tilde u=u,$ $\tilde v=-v,$
$\tilde d=d,$ $\tilde K=-K+1$.
All the above correspondences also mean isomorphisms of
$A^{\max}_{2.*}/\langle\p_v\rangle$ and $A^{\max}_{1}$,
which are realized by means of the projection to the space of $(t,x,u)$ ($\rightarrow$) or
the prolongation on the variable~$v$~($\leftarrow$). Therefore, equation~(\ref{eqf1}) has no
pure potential symmetries for these values of $d$ and $k$.

Cases~2.\ref{t.ps.8z}$^*_{\mu=-4/3}$ and~1.\ref{t58} do not correspond to
each other completely because the basis operator $x^2\p_x-3xu\p_u$ from $A^{\max}_{1.\ref{t58}}$
cannot be prolonged onto $v$ in a local manner
and the algebra $A^{\max}_{2.\ref{t.ps.8z}^*_{\mu=-4/3}}/\langle\p_v\rangle$
is isomorphic to a proper subalgebra of~$A^{\max}_{1.\ref{t58}}$.

\pagebreak

{\begin{center}
{\bf Table 2.} Results of group classification for systems~(\ref{pot1f1})
with respect to $G^{\Equiv}_{{\rm triv.}\pot}$-equivalence
\\[0.5ex] \footnotesize
\setcounter{tbn}{-1}
\renewcommand{\arraystretch}{1.2}
\begin{tabular}{|l|c|c|l|}
\hline\vspacebefore
N &$d(u)$ & $K(u)$ & \hfill {Basis of $A^{\max}$\hfill} \\
\hline\vspacebefore
\refstepcounter{tbn}\thetbn$^*$\label{t.ps.0z}& 
$\forall$&$\forall$&$\p_t,\; \p_x,\; \p_v$\\
\refstepcounter{tbn}\thetbn$^*$\label{t.ps.1z}& 
$\forall$&$0$&$\p_t,\; \p_x,\; \p_v,\; 2t\p_t+x\p_x+v\p_v$\\
\refstepcounter{tbn}\thetbn$^*$\label{t.ps.2z}& 
$e^{\mu u}$&$e^u$&$\p_t,\; \p_x,\; \p_v,\; (\mu-2)t\p_t+(\mu-1)x\p_x+\p_u+((\mu-1)v+x)\p_v$\\
\refstepcounter{tbn}\thetbn$^*$\label{t.ps.3z}& 
$e^u$&$u^2$&$\p_t,\; \p_x,\; \p_v,\; t\p_t+(x+2t)\p_x+\p_u+(x+v)\p_v$\\
\refstepcounter{tbn}\thetbn$^*$\label{t.ps.4z}& 
$e^u$&$0$&$\p_t,\; \p_x,\; \p_v,\; 2t\p_t+x\p_x+v\p_v,\; t\p_t-\p_u-x\p_v$\\
\refstepcounter{tbn}\thetbn$^*$\label{t.ps.5z}& 
$u^{\mu}$&$u^{\nu+1}$&$\p_t,\; \p_x,\; \p_v,\;
(\mu-2\nu)t\p_t+(\mu-\nu)x\p_x+u\p_u+(\mu-\nu+1)v\p_v$\\
\refstepcounter{tbn}\thetbn$^*$\label{t.ps.6z}& 
$u^{\mu}$&$\ln u$&$\p_t,\; \p_x,\; \p_v,\; (\mu+2)t\p_t+(\mu+1)x\p_x+u\p_u+((\mu+2)v-t)\p_v$\\
\refstepcounter{tbn}\thetbn$^*$\label{t.ps.7z}& 
$u^{\mu}$&$u\ln u$& $\p_t,\; \p_x,\; \p_v,\; \mu t\p_t+(\mu x+t)\p_x+u\p_u+(\mu+1)v\p_v$\\
\refstepcounter{tbn}\thetbn$^*$\label{t.ps.8z}& 
$u^{\mu}$&$0$&$\p_t,\; \p_x,\; \p_v,\; 2t\p_t+x\p_x+v\p_v,\; \mu t\p_t-u\p_u-v\p_v$
\setcounter{tbn}{0}\\
\hline
\refstepcounter{tbn}\thetbn\label{t.ps.1}&$u^{-2}e^{\mu/u}$& $ue^{1/u}$&
$\p_t,\; \p_x,\; \p_v,\; (\mu-2)t\p_t+((\mu-1)x+v)\p_x-u^2\p_u+(\mu-1)v\p_v$
\\ 
\refstepcounter{tbn}\thetbn\label{t.ps.2}&$u^{-2}e^{1/u}$& $u^{-1}$&
$\p_t,\; \p_x,\; \p_v,\; t\p_t+(x+v)\p_x-u^2\p_u+(v-2t)\p_v$
\\ 
\refstepcounter{tbn}\thetbn\label{t.ps.3}&$u^{-2}e^{1/u}$& 0&
$\p_t,\; \p_x,\; \p_v,\; 2t\p_t+x\p_x+v\p_v,\; t\p_t-v\p_x+u^2\p_u$
\\
\refstepcounter{tbn}\thetbn\label{t.ps.4}&
$\dfrac {u^\mu}{(u+1)^{\mu+2}}$& $\dfrac {u^{\nu+1}}{(u+1)^{\nu}}$ &
{\arraycolsep=0ex $\begin{array}{l}
\p_t,\; \p_x,\; \p_v,\; (\mu-2\nu)t\p_t+((\mu-\nu)x-v)\p_x+u(u+1)\p_u+\\
+(\mu-\nu+1)v\p_v
\end{array}$}
\\[2ex]
\refstepcounter{tbn}\thetbn\label{t.ps.5}&$\dfrac {u^\mu}{(u+1)^{\mu+2}}$&
$u\ln\dfrac{u}{u+1}$ & $\p_t,\; \p_x,\; \p_v,$ $\mu t\p_t+(\mu x+v-t)\p_x+u(u+1)\p_u+(\mu+1)v\p_v$
\\[2ex]
\refstepcounter{tbn}\thetbn\label{t.ps.6}&$\dfrac {u^\mu}{(u+1)^{\mu+2}}$& $0$ &
$\p_t,\; \p_x,\; \p_v,\; 2t\p_t+x\p_x+v\p_v,\; \mu t\p_t+v\p_x-u(u+1)\p_u-v\p_v$
\\[2ex]
\refstepcounter{tbn}\thetbn\label{t.ps.7}&
$\dfrac{e^{\mu\arctan u}}{u^2+1}$& $\sqrt{u^2+1}\,e^{\nu\arctan u}$ &
{\arraycolsep=0ex $\begin{array}{l}
\p_t,\; \p_x,\; \p_v,\;
(\mu-2\nu)t\p_t+((\mu-\nu)x-v)\p_x+(u^2+1)\p_u+\\+(x+(\mu-\nu)v)\p_v
\end{array}$}
\\
\refstepcounter{tbn}\thetbn\label{t.ps.8}&$\dfrac{e^{\mu\arctan u}}{u^2+1}$& $0$ &
$\p_t,\; \p_x,\; \p_v,\;2t\p_t+x\p_x+v\p_v,\; \mu t\p_t+v\p_x-(u^2+1)\p_u-x\p_v$
\\[1.5ex]
\hline
\refstepcounter{tbn}\thetbn\label{t.ps.9}&$u^{-2}$& $0$ &
$\p_t,\;$ $\p_v,\;$ $2t\p_t+u\p_u+v\p_v,\;$ $-vx\p_x+u(ux+v)\p_u+2t\p_v,$\\
&&&$4t^2\p_t-(v^2+2t)x\p_x+u(v^2+6t+2xuv)\p_u+4tv\p_v,$\\
&&&$x\p_x-u\p_u,\;$ $\phi\p_x-\phi_vu^2\p_u$
\\[0.5ex]
\refstepcounter{tbn}\thetbn\label{t.ps.10}&$u^{-2}$& $u^{-1}$ &
$\p_t,\;$ $\p_v,\;$ $2t\p_t+u\p_u+v\p_v,\;$ $-v\p_x+u^2\p_u+2t\p_v,$\\
&&& $4t^2\p_t-(v^2+2t)\p_x+2u(uv+2t)\p_u+4tv\p_v,$\\
&&& $\p_x,\;$ $e^{-x}\phi\p_x+e^{-x}(\phi-u\phi_v)u\p_u$
\\[0.5ex]
\refstepcounter{tbn}\thetbn\label{t.ps.11}&1&$-u^2$&
$\p_t,\; \p_x,\; 2t\p_t+x\p_x-u\p_u,\; 2t\p_x-\p_u-x\p_v,$\\
&&&$4t^2\p_t+4tx\p_x-2(x+2ut)\p_u-(x^2+2t)\p_v,$\\
&&&$\p_v,\;$ $e^{-v}(h_x-hu)\p_u+e^{-v}h\p_v$
\\[0.5ex]
\refstepcounter{tbn}\thetbn\label{t.ps.12}&1&$0$& $\p_t,\; \p_x,\; 2t\p_t+x\p_x-u\p_u,\;
2t\p_x-(xu+v)\p_u-xv\p_v,\; $\\
&&&$4t^2\p_t+4tx\p_x-((x^2+6t)u+2xv)\p_u-(x^2+2t)v\p_v,$\\
&&&$u\p_u+v\p_v,\; h_x\p_u+h\p_v$\\
\hline
\end{tabular}
\end{center}}

{\footnotesize\noindent
Here $\mu,\nu=\const$. $(\mu,\nu)\not=(-2,-2),\,(0,1)$ and $\nu\not=-1,0$
for Cases~2.\ref{t.ps.5z}$^*$ and~2.\ref{t.ps.4}.
$\mu\not=-2,0$ for Cases~2.\ref{t.ps.8z}$^*$ and~2.\ref{t.ps.6}.
The functions $\phi=\phi(t,v)$ and $h=h(t,x)$ are arbitrary solutions
of the linear heat equation ($\phi_t=\phi_{vv};$ $h_t=h_{xx}$).
\par}

\vspace{2ex}

There are pairs of ``starred" cases from Table~2 which are equivalent with respect to
the hodograph transformation~(\ref{pothodograph}):
2.\ref{t.ps.5z}$^*_{\mu,\nu}$$\leftrightarrow$2.\ref{t.ps.5z}$^*_{\mu',\nu'}$
($\mu+\mu'=-2,$ $\nu+\nu'=1$),
2.\ref{t.ps.6z}$^*_{\mu}$$\leftrightarrow$2.\ref{t.ps.7z}$^*_{\mu'}$,
2.\ref{t.ps.8z}$^*_{\mu}$$\leftrightarrow$2.\ref{t.ps.8z}$^*_{\mu'}$
($\mu+\mu'=-2$).
(To exclude from consideration cases which are equivalent to other
with respect to $G^{\Equiv}_{\pot}$,
we have to assume additionally that e.g. $\mu\ge-1$ and $\nu\ge\frac12$ if $\mu=-1$
for Cases~2.\ref{t.ps.5z}$^*_{\mu,\nu}$ and~2.\ref{t.ps.8z}$^*_{\mu}$.)
Therefore, the following statement is true.

\begin{lemma}
Cases in the pairs
(1.\ref{t56}$_{\mu,\nu}$,1.\ref{t56}$_{\mu',\nu'}$) ($\mu+\mu'=-2,$ $\nu+\nu'=1$),
(1.\ref{t56}$_{\nu=-1}$,1.\ref{t59}),
(1.\ref{t57}a$_{\mu}$,1.\ref{t57}a$_{\mu'}$)
($\mu+\mu'=-2$)
are equivalent with respect to PET~\eqref{pothodograph}.
\end{lemma}

The algebras $A^{\max}_{2.\ref{t.ps.1}}$--$A^{\max}_{2.\ref{t.ps.12}}$ contain operators
which are not projectible to the space~$(t,x,u)$, i.e. their coefficients corresponding to
the variables $t,$ $x$ and $u$ depend on $v$.
Therefore, equation~(\ref{eqf1}) for these values of $d$ and $K$ has purely
potential symmetries.

Cases~2.\ref{t.ps.1}--2.\ref{t.ps.6} including the corresponding Lie invariance algebras
are reduced to ``starred" cases by means of using purely PETs~(\ref{pothodograph}) and~(\ref{pottr}):
2.\ref{t.ps.1}$\rightarrow$2.\ref{t.ps.2z}$^*$,
2.\ref{t.ps.2}$\rightarrow$2.\ref{t.ps.3z}$^*$,
2.\ref{t.ps.3}$\rightarrow$2.\ref{t.ps.4z}$^*$
(hodograph transformation~(\ref{pothodograph}));
2.\ref{t.ps.4}$\rightarrow$2.\ref{t.ps.5z}$^*$,
2.\ref{t.ps.5}$\rightarrow$2.\ref{t.ps.6z}$^*$,
2.\ref{t.ps.1}$\rightarrow$2.\ref{t.ps.2z}$^*$
(transformation~(\ref{pottr}) with $\varepsilon=1$).

Using the equivalence transformation
$\tilde t=t,$ $\tilde x=x-t,$ $\tilde u=u,$ $\tilde v=v+2t,$ $d=d,$ $\tilde K=K-u-2$
from~$G^{\Equiv}_{{\rm triv.}\pot}$,
one can reduce the function $K$ in Case~2.\ref{t.ps.4} ($\nu=-2$) to the simpler form $\tilde K=u^{-1}$.
Then the last operator has the form
$(\mu+4)t\p_t+((\mu+2)x-v)\p_x+u(u+1)\p_u+((\mu+3)v+2t)\p_v$.
A similar statement is true for $\nu=-3$.

\looseness=-1
Cases~2.\ref{t.ps.7} and 2.\ref{t.ps.8} are most specific in the sense of
reducibility to cases from Table~1. There exist no transformations over the real field
that reduce these cases to a simpler form. After considering equation~(\ref{eqf1}) and
system~(\ref{pot1f1}) over the complex field we can reduce Cases~2.\ref{t.ps.7}/2.\ref{t.ps.8}
to Cases 2.\ref{t.ps.5z}$^*_{\mu',\nu'}$/2.\ref{t.ps.8z}$^*_{\mu'}$
where $\mu'=-i\mu/2-1,$ $\nu'=-i\nu/2-1/2$
using the partial case of transformation~(\ref{pottr}):
\[
\tilde t=-4t, \quad \tilde x=-2x+2iv, \quad \tilde v=2x+2iv, \quad
\tilde u=\frac{u-i}{u+i},\quad \tilde d=(u+i)^2d, \quad \tilde K=\frac{K}{u+i}.
\]
Cases~2.\ref{t.ps.7}$_{\mu,\nu}$ and 2.\ref{t.ps.7}$_{\mu',\nu'}$
(2.\ref{t.ps.8}$_{\mu}$ and 2.\ref{t.ps.8}$_{\mu'}$) are equivalent iff
$\mu=-\mu'$ and $\nu=-\nu'$ ($\mu=-\mu'$). The~equivalence is realized by means of
the transformation of changing signs of $x$ and $u$ simultaneously.
To exclude from consideration cases which are equivalent to other
with respect to $G^{\Equiv}_{\pot}$, e.g.,
we have to assume additionally $\mu\ge0$ and $\nu\ge0$ if $\mu=0$
($\mu\ge0$).

\looseness=-1
The $u^{-2}$-diffusion, Fokas--Yortsos, Burgers and linear heat equations
(Cases~1.\ref{t57}a$_{\mu=-2}$, 1.\ref{t57}b, 1.\ref{t510} and 1.\ref{t511} correspondingly)
essentially are distinguished by the group classification of equations~(\ref{eqf1})
in different ways.
After introducing the potential $v$ and replacing equations~(\ref{eqf1}) with
systems~(\ref{pot1f1}) (Cases~2.\ref{t.ps.9}--2.\ref{t.ps.12}), distinction between
these cases and the others becomes more explicit because
$A^{\max}_{2.\ref{t.ps.9}}$--$A^{\max}_{2.\ref{t.ps.12}}$ are isomorphic
infinite-dimensional algebras and,
using PET~(\ref{pothodograph}) and additional equivalence transformation~(\ref{AdEqTr}),
we can transform these cases to each other (see also~\cite{LisleDissertation}):

{\noindent\unitlength=1mm\fboxsep=2mm
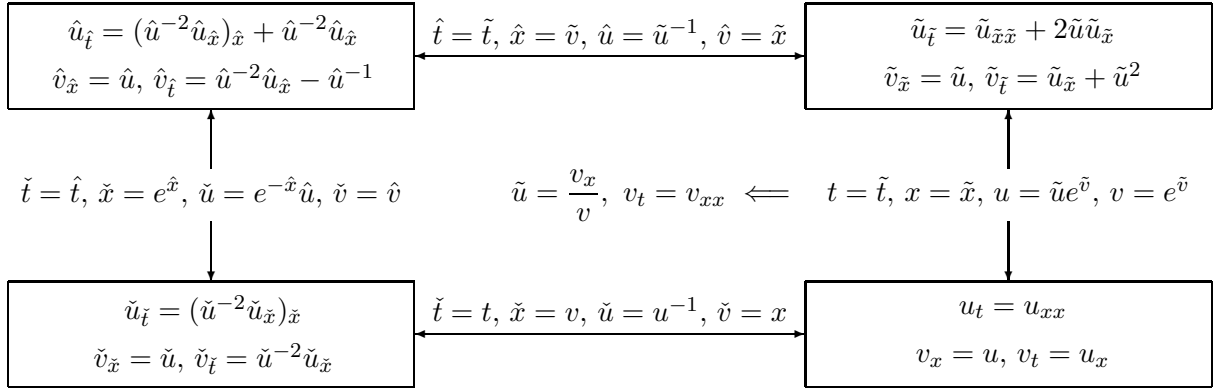
\begin{picture}(160,58)
\put(0,2){\line(0,1){14}}\put(0,2){\line(1,0){54}}\put(54,2){\line(0,1){14}}\put(0,16){\line(1,0){54}}
\put(0,8){{
\parbox{52mm}{\hfil$\check u_{\check t}=(\check u^{-2}\check u_{\check x})_{\check x}$\hfil \\[1ex]
\mbox{}\hfil$\check v_{\check x}=\check u,$
$\check v_{\check t}=\check u^{-2}\check u_{\check x}$\hfil } }}
\put(0,39){\line(0,1){14}}\put(0,39){\line(1,0){54}}
\put(54,39){\line(0,1){14}}\put(0,53){\line(1,0){54}}
\put(0,45){{
\parbox{52mm}{\hfil$\hat u_{\hat t}=
(\hat u^{-2}\hat u_{\hat x})_{\hat x}+\hat u^{-2}\hat u_{\hat x}$\hfil \\[1ex]
\mbox{}\hfil$\hat v_{\hat x}=\hat u,$ $\hat v_{\hat t}=\hat u^{-2}\hat u_{\hat x}-\hat u^{-1}$\hfil } }}
\put(106,39){\line(0,1){14}}\put(106,39){\line(1,0){54}}
\put(160,39){\line(0,1){14}}\put(106,53){\line(1,0){54}}
\put(106,45){ {
\parbox{50mm}{ \hfil$\tilde u_{\tilde t}=
\tilde u_{\tilde x\tilde x}+2\tilde u\tilde u_{\tilde x}$\hfil \\[1ex]
\mbox{}\hfil$\tilde v_{\tilde x}=\tilde u,$
$\tilde v_{\tilde t}=\tilde u_{\tilde x}+\tilde u^2$\hfil } }}
\put(106,2){\line(0,1){14}}\put(106,2){\line(1,0){54}}
\put(160,2){\line(0,1){14}}\put(106,16){\line(1,0){54}}
\put(106,8){ {
\parbox{50mm}{ \hfil$u_t=u_{xx}$\hfil \\[1ex]
\mbox{}\hfil$v_x=u,$ $v_t= u_x$\hfil } }}
\put(54,9){\vector(1,0){52}}\put(106,9){\vector(-1,0){52}}
\put(54,11){\parbox{52mm}{\hfil $\check t=t,$ $\check x=v,$ $\check u=u^{-1},$ $\check v=x$\hfil }}
\put(54,46){\vector(1,0){52}}\put(106,46){\vector(-1,0){52}}
\put(54,48){\parbox{52mm}{\hfil $\hat t=\tilde t,$ $\hat x=\tilde v,$ $\hat u=\tilde u^{-1},$
$\hat v=\tilde x$\hfil }}
\put(27,24){\vector(0,-1){8}}\put(27,31){\vector(0,1){8}}
\put(0,27){\parbox{54mm}{\hfil $\check t=\hat t,$ $\check x=e^{\hat x},$
$\check u=e^{-\hat x}\hat u,$ $\check v=\hat v$\hfil }}
\put(133,24){\vector(0,-1){8}}\put(133,31){\vector(0,1){8}}
\put(106,27){\parbox{54mm}{\hfil $t=\tilde t,$ $x=\tilde x,$ $u=\tilde ue^{\tilde v},$
$v=e^{\tilde v}$\hfil }}
\put(54,26.55){\parbox{52mm}{\hfill $\displaystyle\tilde u=\frac{v_x}v,
\ v_t=v_{xx}\ \Longleftarrow$\hspace*{2mm} }}
\end{picture}}

\vspace{2ex}

Therefore, all equations~(\ref{eqf1}) having infinite-dimensional algebras of
potential symmetries are either linear or linearizable.
As one can see, the well-known Cole--Hopf transformation can be obtained as a combination of
the above transformations.
Only for the linear heat equation the potential symmetry algebra factorized
with $\langle\p_v\rangle$ is isomorphic to the maximal Lie invariance algebra.
The isomorphism is not established with simple projection to the space $(t,x,u).$
It is possible because of linearity and coincidence of the initial and potential equations.

The above analysis results in the following theorem.

\begin{theorem}
All the symmetries presented in Table~2 can be obtained
from Lie symmetries of~\eqref{eqf1} by means of prolongation to the potential $v$
and application of PETs~\eqref{pottr} and~\eqref{pothodograph} (over the complex field in
Cases~2.\ref{t.ps.7} and~2.\ref{t.ps.8})
and additional equivalence
transformation~\eqref{AdEqTr} prolonged to $v$ ($\tilde v =v$).
\end{theorem}

Symmetry properties of systems~\eqref{pot1f1} are connected in a more direct way
with those of~(\ref{pot1f1a}) than with those of~(\ref{eqf1}) because systems~(\ref{pot1f1})
are simply the first prolongation~\cite{Ovsiannikov1} of~(\ref{pot1f1a}) with respect
to the variable~$x$.
Using this connection, we can easily solve the problem of group classification
in the class of equations~(\ref{pot1f1a}).

\begin{theorem}
The equivalence group $\tilde G^{\Equiv}_{\pot}$of the class of equations~\eqref{pot1f1a}
and its Lie algebra~$\tilde A^{\Equiv}_{\pot}$ are projections
of~$G^{\Equiv}_{\pot}$and~$A^{\Equiv}_{\pot}$to the space~$(t,x,v)$.
The Lie algebra of the kernel of principal groups of~\eqref{pot1f1a} is
$\tilde A^{\ker}_{\pot}=\langle \p_t,\; \p_x,\; \p_v\rangle$.
A~complete set of $\tilde G^{\Equiv}_{\pot}$-inequivalent equations~\eqref{pot1f1a}
with the maximal Lie invariance algebra $A^{\max}$ not equal to
$\tilde A^{\ker}_{\pot}$ is exhausted by Cases~2.\ref{t.ps.0z}$^*$--2.\ref{t.ps.4z}$^*$,
2.\ref{t.ps.5z}$^*_{\mu,\nu}$~($\mu\ge-1$ and $\nu\ge\frac12$ if $\mu=-1$),
2.\ref{t.ps.6z}$^*_{\mu}$, 2.\ref{t.ps.8z}$^*_{\mu}$~($\mu\ge-1$),
2.\ref{t.ps.7}~($\mu\ge0$ and $\nu\ge0$ if $\mu=0$),
2.\ref{t.ps.8}~($\mu\ge0$) and 2.\ref{t.ps.12}.
\end{theorem}

\section{PETs as nonlocal symmetry transformations}

There exist equations in class~(\ref{eqf1}) that are invariant with respect to
nontrivial transformations from~$G^{\Equiv}_{\pot}$. Thus, equation~(\ref{eqf1})
admits transformations either~(\ref{pottr},$\varepsilon=1$) or~(\ref{pothodograph}) iff
\[
\mbox{either} \quad
d=u^{-2}F^1(u^{-1}), \; K=uG^1(u^{-1})
\quad \mbox
{or} \quad
d=u^{-1}F^2(\ln u), \; K=u^{1/2}G^2(\ln u),
\]
where $F^1$ and $G^1$ are periodic functions with the period equal to 1 and
$F^2$ ($G^2$) is an even (odd) function.
Such nonlocal symmetry transformations generate additional
(with respect to Lie symmetries) equivalences in sets of solutions.

Consider, in more detail, the fast diffusion equation
\begin{equation}\label{fde}
u_t=(u^{-1}u_x)_x.
\end{equation}
It is invariant with respect to transformation~\eqref{pothodograph}
which is additional to the usual Lie symmetry group~$G^{\max}$ of equation~\eqref{fde}.
Action of elements of~$G^{\max}$ on the solutions is given by the formula~\cite{Ovsiannikov1959}
\[
\tilde u(t,x)=\varepsilon_3^{-1}\varepsilon_4^2\,u(\varepsilon_3t+\varepsilon_1,\varepsilon_4x+\varepsilon_2),
\]
where $\varepsilon_1$, \dots, $\varepsilon_4$ are arbitrary constants,
$\varepsilon_3\varepsilon_4\ne0$.

\begin{lemma}
The set of Lie invariant solutions of equation~\eqref{fde} is closed under transformation~\eqref{pothodograph}.
\end{lemma}
\begin{proof}
Transformation~\eqref{pothodograph} generates an adjoint action~$\mathcal{H}$ on the Lie symmetry algebra 
\[
A^{\max}_{\pot}=\langle \p_t,\; \p_x,\; \p_v,\;\hat D^1=x\p_x-2u\p_u-v\p_v,\;\hat D^2=2t\p_t+x\p_x+v\p_v\rangle
\]
of the corresponding potential system (Case~2.\ref{t.ps.8z}$^*_{\mu=-1}$), which is determined in the following way:
$\mathcal{H}(\p_t)=\p_t$,
$\mathcal{H}(\p_x)=\p_v$,
$\mathcal{H}(\p_v)=\p_x$,
$\mathcal{H}(\hat D^1)=-\hat D^1$,
$\mathcal{H}(\hat D^2)=\hat D^2$.
Elements of the Lie symmetry algebra 
\[
A^{\max}=\langle \p_t,\; \p_x,\; D^1=x\p_x-2u\p_u,\; D^2=2t\p_t+x\p_x\rangle
\]
of equation~\eqref{fde} is prolonged with respect to~$v$ ambiguously up to a term proportional to~$\p_v$. 
Therefore, transformation~\eqref{pothodograph} correctly generates also an adjoint action~$\mathcal{H}'$ on the classes 
of elements from~$A^{\max}$, which differ each from other with terms proportional to~$\p_v$.

In view of the above-mentioned, up to translations with respect to~$v$ (or~$x$) 
any invariant solution of~\eqref{fde} gives 
an invariant solution of the potential system, 
which is transformed by~\eqref{pothodograph} to an invariant solution of the same system, and 
the latter solution can be projected to an invariant solution of~\eqref{fde}.
\end{proof}

All invariant solutions constructed in closed forms earlier with the classical Lie method 
were collected e.g. in~\cite{Polyanin&Zaitsev}. 
A complete list of $G^{\max}$-inequivalent solutions of such type are exhausted by the following ones:
\begin{gather}\label{Liesolutions.for.u-1}
\begin{split}&
1)\ u=\dfrac{1}{1+\varepsilon e^{x+t}};\quad 
2)\ u=e^{x};\quad 
3)\ u=\dfrac{1}{x-t+\mu te^{-x/t}};
\\[1ex]&
4)\ u=\dfrac{2t}{x^2+\varepsilon t^2};\quad
5)\ u=\dfrac{2t}{\cos^2x};\quad 
6)\ u=\dfrac{-2t}{\cosh^2x};\quad
7)\ u=\dfrac{2t}{\sinh^2x}.
\end{split}\end{gather}
Here $\varepsilon$ and $\mu$ are arbitrary constants, $\varepsilon\in\{-1,0,1\}\!\!\mod G^{\max}\!$.
The below arrows denote the possible transformations of solutions~\eqref{Liesolutions.for.u-1}
to each other by means of~\eqref{pothodograph} up to translations with respect to~$x$:
\[
\begin{split}&
\mbox{\Large$\circlearrowright$}\;1)_{\varepsilon=0}\,; \quad
1)_{\varepsilon=1}\longleftrightarrow 1)_{\varepsilon=-1,\;x+t<0}\,; \quad
\mbox{\Large$\circlearrowright$}\;1)_{\varepsilon=-1,\;x+t>0}\,; \quad
2)\longleftrightarrow 3)_{\mu=0,\;x>t}\,; \\&
\mbox{\Large$\circlearrowright$}\;4)_{\varepsilon=0}\,; \quad
5)\longleftrightarrow 4)_{\varepsilon=4}\,; \quad
6)\longleftrightarrow 4)_{\varepsilon=-4,\;|x|<2|t|}\,; \quad
7)\longleftrightarrow 4)_{\varepsilon=-4,\;|x|>2|t|}\,. 
\end{split}
\]
The sixth connection was known earlier~\cite{Fushchych&Serov&Amerov1982,Pukhnachov1996}.
If $\mu\ne0$ solution 3) from list~\eqref{Liesolutions.for.u-1} is mapped by~\eqref{pothodograph} 
to the solution 
\[
8)\ u=t\vartheta(\omega)-t+\mu te^{-\vartheta(\omega)},\qquad \omega=x-\ln|t|,
\]
which is invariant with respect to the algebra~$\langle t\p_t+\p_x+u\p_u\rangle$.
Here $\vartheta$ is the function determined implicitly by the formula
$
\int (\vartheta-1+\mu e^{-\vartheta})^{-1}d\vartheta=\omega.
$ 

To find exact solutions of equation~\eqref{fde}, other methods can be used also. 
Thus, M.L.~Gandarias~\cite{Gandarias2001} found the new exact non-Lie solutions with the nonclassical symmetry 
method. We adduce the list of these solutions up to $G^{\max}$-equivalence, completing it with similar ones:
\begin{equation}\label{nonLiesolutions.for.u-1}
\begin{split}&
1)\ u=\dfrac{\cos t}{\sin x-\sin t};\quad 
2)\ u=\dfrac{\cosh t}{\sinh x-\sinh t};\quad 
3)\ u=\dfrac{-\sinh t}{\cosh x+\cosh t};
\\[1ex]&
4)\ u=\dfrac{\sinh t}{\cosh x-\cosh t};\quad
5)\ u=\dfrac{\cos t}{\cosh x+\sin t};\quad
6)\ u=\dfrac{\sinh t}{\cosh x+\cosh t}.
\end{split}\end{equation}
Solutions~\eqref{nonLiesolutions.for.u-1} can be presented in the form of compositions of two simple waves 
which move with the same velocities in opposite directions:  
\[\begin{split}&
1)\ u=\cot(x-t)+\tan(x+t);\quad 
2)\ u=\coth(x-t)-\tanh(x+t);\\[1ex]& 
3)\ u=\tanh(x-t)-\tanh(x+t);\quad
4)\ u=\coth(x+t)-\coth(x-t).
\end{split}\]
(We simplify the above representations by a scale transformation. 
The fifth and sixth solutions admit such representation over the complex field only.)
Solutions~4), 5) and~6) are not adduced in~\cite{Gandarias2001} in any form. 

Up to translations with respect to~$x$,
transformation~(\ref{pothodograph}) acts on the set of solutions~\eqref{nonLiesolutions.for.u-1} 
in the following way:
\[
\begin{split}&
1)_{|\sin x|>|\sin t|} \longleftrightarrow 5)\,; \quad
1)_{|\sin x|<|\sin t|} \longleftrightarrow 5)|_{x\to-x}\,;
\\&
\mbox{\Large$\circlearrowright$}\;2)_{x<t}\,; \quad
2)_{x>t} \longleftrightarrow 2)_{x>t}|_{x\to-x}\, ;\quad
3)\longleftrightarrow 4)_{|x|<|t|}\,; \quad
\mbox{\Large$\circlearrowright$}\;4)_{|x|>|t|}\,; \quad
\mbox{\Large$\circlearrowright$}\;6).
\end{split}
\]
The latter actions can be interpreted in terms of actions of transformation~(\ref{pothodograph}) on 
the nonclassical symmetry operators which correspond to solutions~\eqref{nonLiesolutions.for.u-1}.

\section{Conclusion}

It was proved above that any nonlinear diffusion--convection equation having non-trivial potential symmetries
can be reduced to another diffusion--convection equation with potential equivalence transformations
such that all symmetries will become point. This result generates a number of questions, and each from them is
an interesting problem. Are similar statements right for more general classes of differential equations?
Does an differential equation exist, potential symmetries of which cannot be constructed from point symmetries of
an equation equivalent to the initial one via potential transformations?
Could the result be generalized to other kinds of symmetries (e.g. nonclassical, conditional and approximate ones)?
We believe that solving the above problems allows us to understand deeper the essence of potential symmetries.

We have also investigated the other potential forms of equations~(\ref{eqf1}).
These results will be published in our further paper on the subject.

\subsection*{Acknowledgements}

The authors are grateful to
Profs. V.~Boyko, A.~Nikitin, O.~Morozov, C.~Sophocleous, I.~Yehor\-chenko and A.~Zhalij
for useful discussions and interesting comments.
The research of NMI was partially supported by National Academy of Science of Ukraine
in the form of the grant for young scientists.
ROP thanks Prof.~F.~Ardalan and Dr.~H.~Eshraghi
(School of Physics, Institute for Studies in Theoretical Physics and Mathematics, Tehran, Iran)
and Prof.~J.~Patera (Centre de recherches math{\'e}matiques,
Universit{\'e} de Montr{\'e}al, Canada)
for hospitality and support during writing this paper.

\end{document}